\documentclass[a4paper,prd,preprint,amsmath,
       superscriptaddress,nofootinbib,floatfix,showpacs]{revtex4}
\usepackage{graphicx}
\renewcommand{\theequation}{\arabic{equation}}
\renewcommand{\thesection}{\arabic{section}}
\renewcommand{\thefootnote}{\fnsymbol{footnote}}

\newcommand{\vs}[1]{\vspace{#1 mm}}

\renewcommand{\a}{\alpha}
\renewcommand{\b}{\beta}

\newcommand{\e}{\epsilon}

\begin{document}
\noindent
\topmargin 0pt
\oddsidemargin 5mm

\begin{titlepage}
\setcounter{page}{0}
\thispagestyle{empty}
\begin{flushright}
June, 2002\\
OU-HET 414\\
hep-ph/0206207\\
\end{flushright}
\vs{4}
\begin{center}
{\LARGE{\bf Exploring the neutrino mass matrix at 
$M_R$ scale }} \\ 
\vs{6}
Takahiro Miura\footnote{e-mail address:
miura@het.phys.sci.osaka-u.ac.jp},
Tetsuo Shindou\footnote{e-mail address:
shindou@het.phys.sci.osaka-u.ac.jp} and 
Eiichi Takasugi\footnote{e-mail address:
takasugi@het.phys.sci.osaka-u.ac.jp}\\
\vs{2}
{\em Department of Physics,
Osaka University \\ Toyonaka, Osaka 560-0043, Japan} \\
\end{center}
\vs{6}
\centerline{{\bf Abstract}}
We discuss the neutrino mass matrix which predicts zero or 
small values of $|V_{13}|$ in the MSSM and 
found the inequality, $\sin^2 2\theta_{12}\le \sin^2 2\theta_{\odot}$, 
where $\sin^2 2\theta_{12}$ is the mixing angle at $M_R$ scale and 
$\sin^2 2\theta_{\odot}$ is the value determined by the solar 
neutrino oscillation. This constraint says 
that the model which predicts a larger 
value of $\tan^2 \theta_{12}$ at $M_R$ 
than the experimental value is excluded. In particular, the bi-maximal 
mixing scheme at $M_R$ scale is excluded, from the experimental 
value $\tan^2 \theta_\odot<1$. In this model, $|V_{13}|$  and 
a Dirac phase at $m_Z$ which are induced radiatively may not be 
small. 

\end{titlepage}

\newpage
\renewcommand{\thefootnote}{\arabic{footnote}}
\setcounter{footnote}{0}

\section{Introduction}

The SuperKamiokande data has discovered the neutrino 
mixing between $\nu_\mu$ and $\nu_\tau$ from the 
atmospheric data\cite{SuperK-Atm}. Now the SNO data\cite{SNO-Solar} 
together with 
SuperKamiokande data\cite{SK-Solar} have solved the solar neutrino 
puzzle and pinpointed a solution among 
four solutions. That is, its origin is 
mainly due to the mixing between the $\nu_e$ and $\nu_\mu$ 
and the LMA solution is the most probable one. 
They are 
summarized as
\begin{eqnarray}
\tan^2 \theta_\odot\simeq 0.34\;,&&
\Delta m_\odot^2\simeq 5\times 10^{-5}\;{\rm eV}^2\;, 
\nonumber\\ 
\sin^22\theta_{\rm atm}\simeq 1\;,&&
\Delta m_{\rm atm}^2 \simeq 3\times 10^{-3}\;{\rm eV}^2\;,
\end{eqnarray}
where $\theta_\odot$ and $\theta_{\rm atm}$ are mixing angles 
which appear in the solar and atmospheric neutrino 
oscillations, and in effect are mixing angles 
between the 1st and the 2nd and between the 2nd and the 3rd 
mass eigenstates, respectively. $\Delta m_\odot^2$
and $\Delta m_{\rm atm}^2$ are the squared mass differences defined by 
$\Delta m_\odot^2=m_2^2-m_1^2$ and $ \Delta m_{\rm
atm}^2=|m_3^2-m_2^2|$, 
where $m_i$ is the mass of the $i$-th mass eigenstate of neutrinos. 
Usually, the sign convention, $\Delta m_\odot^2=m_2^2-m_1^2>0$ is 
taken, in which case the result from the SNO-SuperKamiolande 
data\cite{SNO-Solar,SK-Solar} favors
the normal side, $\tan^2 \theta_\odot<1$, and disfavors 
the dark side, $\tan^2 \theta_\odot>1$ \cite{Darkside}. 
Another important information from the CHOOZ data\cite{CHOOZ}
gives a severe upper limit for $|V_{13}|$
\begin{eqnarray}
|V_{13}|<0.16\;,
\end{eqnarray}
where $V_{13}$ is the element of the MNS neutrino mixing
matrix\cite{Maki:mu}
representing the mixing between the 1st and the 3rd mass eigenstates. 
If we combine these information, the neutrino mixing
matrix
is approximately written by
\begin{eqnarray}
V=\begin{pmatrix}
c_\odot &-s_\odot &0\cr
s_\odot c_{\rm atm}&c_\odot c_{\rm atm}&-s_{\rm atm}\cr
s_\odot s_{\rm atm}&c_\odot s_{\rm atm}&c_{\rm atm}\cr
\end{pmatrix}
\begin{pmatrix}
1&&\cr
&e^{i\a_M}&\cr
&&e^{i\b_M}\cr
\end{pmatrix}\;,
\end{eqnarray}
where we included two Majorana CP violation 
phases\cite{Bilenky:1980cx,DBD-Majo,SV} in the 
mixing matrix which play an important role for the neutrinoless 
double beta decay\cite{DBD-Majo}. 

Among the above experimental information, the most mysterious 
point is the puzzle why $|V_{13}|$ is so small 
in comparison with other mixing angles,  $\theta_{\odot}$ and 
$\theta_{\rm atm}$. If it is really small, we have to find out 
the reason for it. For a small quantity at the low energy scale, 
the naturalness usually asserts that it is zero at the higher 
energy scale, because it is quite hard to reproduce such a small 
quantity at the low energy scale. 

In this paper, we consider a possibility that $|V_{13}|=0$ 
at the energy scale where the left-handed neutrino mass is 
induced by the see-saw mechanism. There are many advantages 
to consider this possibility. (1) The small value of $|V_{13}|$ 
is naturally explained because it is induced by the radiative 
correction. (2) This scenario may be realized in some theoretical 
models at $M_R$ scale\cite{BiMax,Demo,TriMax,Fukuura:1999ze}. 
(3) The Dirac CP violation phase is induced by the radiative 
correction. 

Now, we consider the neutrino mass matrix which predicts $|V_{13}|=0$ 
at $M_R$ scale, in the diagonal basis of the charged lepton 
mass matrix. This mass matrix contains only seven parameters, 
three neutrino masses, two mixing angles and two Majorana 
CP violation phases, and thus there is no Dirac CP violation 
phase at $M_R$ scale. 
This may gives an possibility that the Dirac CP violation 
phase which appears in the neutrino oscillation may be related to 
Majorana CP violation phases, which may be related 
in the leptogenesis, since in our model, two Majorana phases 
are associated with phases of neutrino masses and they may well 
have some relation with  
phases from the heavy right-handed Majorana mass matrix.

This paper is organized as follows: In Section 2, we explain 
our model and the framework of neutrino mass matrix. The 
 radiative correction is taken into account and the mass matrix 
is diagonalized analytically to connect the parameters at $M_R$ 
scale and the present experimental scale, $m_Z$. In Section 3, 
the general feature of our result is explained and the predictions 
are given. In Section 4, by using analytic result, numerical analysis 
is made on the induced size of $|V_{13}|$, the Dirac CP violation phase, 
$\delta$, and the effective mass of the neutrinoless double beta 
decay. The discussion on the absolute size of neutrino mass is given. 
Summary and discussions are given in Section 5.

\section{The model} 

We consider a class of left-handed neutrino mass matrices 
which gives $V_{13}=0$, where $V$ is the neutrino mixing 
matrix. We assume that this mass matrix 
is derived by the see-saw mechanism according to 
the SUSY GUT scenario at the right-handed neutrino mass, 
$M_R$ scale and evolves following the renormalization 
equation for MSSM to the Z boson mass scale, $m_Z$. 
In this model, $V_{13}$ at $m_Z$ scale is induced by 
radiative correction. We examine the size of $V_{13}$, 
the Dirac CP violation phase, $\delta$, Majorana CP 
violation phases, $\a_M$ and $\b_M$, and the effective 
mass of the neutrinoless double beta decay, $\langle m_\nu\rangle$. 

\noindent
(a) The mass matrix at $M_R$

The mass matrix which gives $V_{13}=0$ is generally expressed 
in the diagonal basis of the charged lepton mass matrix as
\begin{eqnarray}
m_\nu(M_R) =O D_\nu O^T \;,
\end{eqnarray}
where
\begin{eqnarray}
D_\nu={\rm diag}(M_1, M_2e^{i\alpha_0}, M_3e^{i\beta_0})\;,
\end{eqnarray}
with Majorana phases, $\alpha_0$ and $\beta_0$. $O$ is 
the mixing matrix at $M_R$ scale and is given by 
\begin{eqnarray}
O=\begin{pmatrix}
1&0&0\cr
0&c_{23}&-s_{23}\cr
0& s_{23}& c_{23}\cr
\end{pmatrix}
\begin{pmatrix}
 c_{12}&- s_{12}&0\cr
 s_{12}& c_{12}&0\cr
0&0&1\cr\end{pmatrix}
\;.
\end{eqnarray}
with $ c_{ij}=\cos \theta_{ij}$ and $ s_{ij}
=\sin  \theta_{ij}$. 
In the following, we use $\theta_{ij}$ only as an angle 
at $M_R$ scale.

\noindent
(b) The neutrino mass matrix at $m_Z$

In MSSM, the neutrino mass matrix at $m_Z$ is given by\cite{RGE-formula}
\begin{eqnarray}
m_\nu(m_Z)={\rm diag}(1,1,\a) O D_\nu O^T{\rm diag}(1,1,\a)
\;,
\end{eqnarray}
where $\a$ is defined by
\begin{eqnarray}
\a \equiv 1- \epsilon =1/\sqrt{I_{\tau}}
= \Biggl( \frac{m_Z}{M_R} \Biggr)
^{\frac{1}{8\pi^2}(1+\tan^2\b)(m_{\tau}/v)^2} <1\;.
\end{eqnarray}
Here $m_{\tau}$ is the $\tau$ lepton mass,
$v^2=v^2_u +v^2_d$ and $\tan\b = v_u/v_d$ with
$v_i$ being the vacuum expectation value of 
MSSM Higgs doublet $\langle H_i\rangle(i=u,d)$. 

In order to estimate $\epsilon$, we assume 
the right-handed mass scale, $M_R$ and 
the region of $\tan \beta$ as
\begin{eqnarray}
M_{\rm R}=10^{13}({\rm GeV})\;,\qquad 2< \tan \beta <50\;.
\end{eqnarray} 
Then, with  $m_Z=91.187({\rm GeV})$, $m_{\tau}=1.777
({\rm GeV})$ and $v=245.4({\rm GeV})$, we find 
\begin{eqnarray}
8\times 10^{-4}< \epsilon < 5\times 10^{-2}\;.
\end{eqnarray}

\noindent
(c) Masses and the mixing matrix at $m_Z$

The effect of the radiative correction to neutrino mass matrix has 
been discussed by many authors\cite{RGE-formula,Stability,RGEwithMaj} 
and the following is known. 
(1) The mixing angles are stable for the case of the hierarchical 
or the inverted-hierarchical neutrino mass scheme, $m_1\ll m_2\ll m_3$ 
or $m_3\ll m_1\ll m_2$. (2) The instability occurs for $m_1\simeq m_2$. 
(3) The Majorana phases in neutrino masses may play an important 
role\cite{RGEwithMaj}.

Since the stable case is well analyzed, we focus on the unstable 
case. That is, we consider the following mass relation holds 
at $M_R$ scale, 
\begin{align}
&M_1\simeq M_2\;, \;0<\Delta_{21}\ll |\Delta_{31}|\;,
\nonumber\\ 
&0<\Delta_{21}\ll M_1^2\;,
\;\e M_1^2\simeq \e M_2^2\ll|\Delta_{31}|\;,
\end{align}
where $\Delta_{21}=M_2^2-M_1^2$, $\Delta_{31}=M_3^2-M_1^2$ 
and we chose the convention, $\Delta_{21}>0$. 
The diagonalization of the neutrino mass matrix is made 
analytically and the derivation is given in Appendix.  

In the following, we summarize the result derived in 
Appendix. As for neutrino masses themselves, 
corrections are small and of order $\e M_i$,
because we are considering the situation where 
$M_1^2\simeq M_2^2\gg \Delta_{21}\sim \e M_2^2$. 
Thus, neutrino masses at $M_R$ and $m_Z$ can be 
considered to be the same.  
\begin{eqnarray}
m_i \simeq M_i\;.
\end{eqnarray}
The radiative correction gives effect to the mass difference between 
$m_1$ and $m_2$,
\begin{eqnarray}
\Delta m_{\odot}^2 =m_2^2-m_1^2=\frac{\Delta_{21}-
   4\e m_1^2 s_{\rm atm}^2 \cos 2 \theta_{12} }{\cos 2\theta}>0
   \;,
\end{eqnarray}
while the mass difference between the 2nd and the 3rd receives 
only a negligible effect, so that
\begin{eqnarray}
  \Delta m_{\rm atm}^2 =m_3^2-m_2^2
  \simeq m_3^2-m_1^2 \simeq \Delta_{31}\;. 
\end{eqnarray}
In the above, we required $\Delta m_{\odot}^2 >0$ in accordance 
with the common experimental analysis which gives 
$\tan^2 \theta_\odot =0.34 <1$. As for mixing angles, 
the radiative correction 
does not give any effect to the mixing between the 2nd and 
the 3rd mass eigenstates either. That is, 
\begin{eqnarray}
\theta_{\rm atm}=\theta_{23}\;.
\end{eqnarray}
Thus, in the following, we use $m_i$ for $M_i$ except for 
the discussion of the mass difference and $\theta_{\rm atm}$ 
for $\theta_{23}$ in order to express $\theta_\odot$, 
$|V_{13}|$, $\delta$, $\a_M$ and $\b_M$ in terms of observables 
at the low energy as possible as we can. 

The MNS mixing matrix which is given in Eq.~(A.15) is expressed as
\begin{eqnarray}
V=\begin{pmatrix}
c_\odot &-s_\odot &-|V_{13}|e^{-i\delta}\cr
s_\odot c_{\rm atm}&c_\odot c_{\rm atm}&-s_{\rm atm}\cr
s_\odot s_{\rm atm}&c_\odot s_{\rm atm}&c_{\rm atm}\cr
\end{pmatrix}
\begin{pmatrix}
1&&\cr
&e^{i\a_M}&\cr
&&e^{i\b_M}\cr\end{pmatrix}\;,
\end{eqnarray}
where $\theta_{\rm atm}=\theta_{23}$, $\theta_{\odot}$ is 
\begin{eqnarray}
s_{\odot}&=&| s_{12}c+ c_{12}s e^{i\a_0/2}|\;,
\nonumber\\
c_{\odot}&=&| c_{12}c- s_{12}se^{-i\a_0/2}|\;,
\end{eqnarray}
with $\theta$ defined by
\begin{eqnarray}
\sin 2\theta=  \frac{4\epsilon m_1^2 s_{\rm atm}^2
   \sin 2 \theta_{12}\cos \frac{\a_0}2}{\Delta m^2_{\odot}}
  \;.
\end{eqnarray}
The induced mixing element, $|V_{13}|$ is given by
\begin{eqnarray}
|V_{13}|&=&\frac{\epsilon m_1m_3 
\sin 2 \theta_{12}\sin 2\theta_{\rm atm}
\sin\frac{\alpha_0}{2}}
{\Delta m^2_{\rm{atm}}}\;.
\end{eqnarray}
A Dirac CP violation phase, $\delta$, and two Majorana CP violating 
phases, $\a_M$ and $\b_M$ are 
\begin{eqnarray}
\delta&=&\xi_1+\xi_2-\frac{\pi}2+\frac{\a_0}2-\beta_0\;,
\nonumber\\
\a_M&=& \xi_2-\xi_1-\frac{\a_0}2\;,\nonumber\\ 
\b_M&=&\xi_2-\frac{\b_0}2\;,
\end{eqnarray}
with
\begin{eqnarray}
\xi_1&=&\arg( c_{12}c- s_{12}se^{-i\a_0/2})\;,
\nonumber\\
\xi_2&=&\arg( s_{12}c+ c_{12}se^{i\a_0/2})\;.
\end{eqnarray}
In the mixing matrix, $\theta_{12}$ and two Majorana phases, $\a_0$ and 
$\b_0$ are only parameters defined at $M_R$. All other parameters are 
expressed by physical quantities at $m_Z$.

\section{General features}

As we explained, we take the convention $\Delta m_{\odot}^2>0$ for 
which the result from the SNO-SuperKamiokande data 
requires
that the mixing angle should be in the normal side, 
$\tan^2 \theta_\odot =0.34<1$\cite{SNO-Solar,SK-Solar}. 
Also we take the convention, 
$\Delta_{21}=M_2^2-M_1^2>0$. 

\vskip 2mm
\noindent
(a) The solar mixing angle

Here, we discuss the relation between 
$\theta_{12}$ defined at $M_R$ scale and  $\theta_{\odot}$ 
defined at $m_Z$, the value from the solar neutrino 
oscillation data. 
We parametrize the solar neutrino mixing angle as 
\begin{eqnarray}
\tan^2 \theta_{\odot}&=&\frac{1-p}{1+p}\;,
\;\;{\rm or}\;\;
\sin^2 2\theta_{\odot}=1-p^2\;,
\end{eqnarray}
and then $p>0$ is required to guarantee $\tan^2 \theta_{\odot}<1$. 
From Eq.~(17), $p$ is given by 
\begin{eqnarray}
p&=&\cos2 \theta_{12}\cos2 \theta 
-\sin2 \theta_{12}\sin2 \theta\cos\frac{\a_0}2
\nonumber\\
&=&\cos2 \theta_{12}\cos2 \theta 
-2h \sin^2 2 \theta_{12}\cos^2\frac{\a_0}2\;,
\end{eqnarray}
where we used $\sin 2\theta$ defined in Eq.~(18) to derive 
the second line and we defined the positive quantity $h$ 
to avoid the complexity of equation,
\begin{eqnarray}
h=\frac{2\epsilon m_1^2 s_{\rm atm}^2}{\Delta m^2_{\odot}}\;.
\end{eqnarray}
From $p>0$ together with $h>0$, we find 
$\cos2 \theta_{12}\cos2 \theta>0$. Now we look carefully 
the equation for $\Delta m_\odot^2>0$ in Eq.~(13). With 
$\Delta_{21}>0$ together with the above condition, only 
consistent choice is 
\begin{eqnarray}
\cos 2 \theta_{12}>0\;,\;\;\cos 2\theta>0\;.
\end{eqnarray}
Now that the sign of $\cos 2\theta$ is fixed to be positive, 
we can eliminate $\cos 2\theta$ in Eq.~(23). Thus, we can express 
$\tan^2 \theta_{\odot}$ in terms of $\theta_{12}$, 
$\a_0$ and $h$,
\begin{eqnarray}
\frac{1-\tan^2 \theta_{\odot}}{1+\tan^2 \theta_{\odot}}
=\cos 2\theta_{12}\sqrt{1-4h^2\sin^2 2 \theta_{12} 
\cos^2 \frac{\a_0}2}-2h\sin^2 2 \theta_{12} 
\cos^2 \frac{\a_0}2\;.
\end{eqnarray}
This is the equation which relates the angle $\theta_{12}$ at 
$M_R$ scale and the solar mixing angle, $\theta_\odot$. 

Next, we solve the above equation with respect to $\cos\a_0$ 
and find
\begin{eqnarray}
1+\cos\a_0=\frac{1}{h\sin^22 \theta_{12}}\left(-|\cos 2\theta_\odot|
-h\cos^22 \theta_{12}
+ \cos2 \theta_{12}
\sqrt{h^2\cos^22 \theta_{12}+2|\cos 2\theta_\odot h+1}\right)\;.
\end{eqnarray}
By requiring $0\leq 1+\cos\alpha \leq 2$, we obtain
\begin{eqnarray}
\frac{\sin^2 2\theta_{\odot}}{\sin^2 2\theta_{\odot}
+(|\cos 2\theta_{\odot}|
+4\e s_{\rm atm}^2 (m_1^2/\Delta m_{\odot}^2) )^2
}\leq \sin^22 \theta_{12}
\leq \sin^22\theta_{\odot}\;,
\end{eqnarray}
where we used the expression of $h$ in Eq.~(24). 
This inequality shows the region of 
$\sin^2 2 \theta_{12}$ at $M_R$ where  
the experimental value $\sin^22\theta_{\odot}$ at $m_Z$ is 
realized. 

Before discussing the meaning of these equations, it should be 
mentioned that the above result is valid as far as 
$|V_{13}(M_R)|\ll|\sin\theta_{12}|$ is satisfied, where $\sin
\theta_{12}$ 
and $|V_{13}(M_R)|$ are quantities at $M_R$ scale. 

Now we discuss the meaning of the inequality. Firstly, we comment 
that the equality $\sin^22\theta_{12}=\sin^2 2\theta_{\odot}$ 
(stable case) holds in several cases, (1) the hierarchical mass case, 
$m_1\ll m_2$, where $m_2^2\simeq \Delta m_\odot^2$, 
(2) the small $\epsilon$ case which corresponds to 
the small $\tan \beta$, (3) the special case, $\a_0=\pi$, even with  
$m_1^2\simeq m_2^2\gg\Delta m_\odot^2$. 

One of the most important observation will be the following: 
The inequality $\sin^22\theta_{12}\le \sin^2 2\theta_{\odot}$ 
holds for most of models, because physically feasible models 
have to predict small values of $|V_{13}|$. Thus, we can generally 
say that the model which constructed at higher energy scale 
such as $M_R$ must have $\tan^2 \theta_{12}$ less than or equal to 
the experimental value, $\tan^2 \theta_{\odot}=0.34$. 
The bi-maximal mixing scheme which is realized at $M_R$ is not 
acceptable from the present experimental data. 
This may give a big obstacle for model building, because 
the model should predicts the experimental angle which 
does not have any particular meaning in the stable angle 
case. On the other hand, for the unstable case, the model 
needs to predict smaller value at $M_R$ and the radiative 
correction lifts the value to the experimental one, by 
the interplay among neutrino mass, $\tan \beta$ and 
the CP violation angle, $\a_0$. 

\vskip 2mm
\noindent
(b) The size of the induced $|V_{13}|$

The induced  $|V_{13}|$ is given in 
Eq.~(19). Since it is proportional to 
$\epsilon (m_1m_3/\Delta m_{\rm atm}^2)$, its value is 
suppressed by $\Delta m_{\odot}^2/\Delta m_{\rm atm}^2$, in 
comparison with corrections to the mass squared difference for 
the solar neutrino mixing and the solar neutrino angle. 
If $m_3/m_1>1$, some enhancement is expected. 

\vskip 2mm
\noindent
(c) The CP violation angles

The Dirac CP violation phase, $\delta$ is induced from two Majorana 
phases. Since $\a_0$ is deeply involved in determining the solar mixing 
angle, $\delta$ aside from $\b_0$ can be determined. Thus, 
we define
\begin{eqnarray}
\delta_1=\delta+\b_0=\xi_1+\xi_2-\frac{\pi}2+\frac{\a_0}2\;,
\end{eqnarray}
for which we analyze numerically in the next section. We hope that 
the knowledge of the phase $\b_0$ may be derived from the information 
from the leptogenesis. 

\vskip 2mm
\noindent
(d) Neutrinoless double beta decay

With $m_1\simeq m_2$, 
the effective neutrino mass for the neutrinoless double 
beta decay in this mode is simply given by
\begin{eqnarray}
\langle m_{\nu}\rangle
&\equiv& \left|\sum_j m_jV_{1j}^2\right|\nonumber\\
&\simeq& m_1\left|( c_{12}c- s_{12}s e^{-i\a_0/2})^2
+( s_{12}c+ c_{12}se^{i\a_0/2})^2e^{-i\a_0}\right|
\nonumber\\
&=& m_1\sqrt{1-\sin^22
\theta_{12}\sin^2\frac{\alpha_0}{2}}\;,
\end{eqnarray}
where we neglect $m_3V_{13}^2$, because $|V_{13}|$ is 
small.

\section{Numerical analysis}

Physical quantities, $\tan^2\theta_{\odot}$, $\Delta m_{\odot}^2$, 
$\sin^22\theta_{\rm atm}$, $\Delta m_{\rm atm}^2$, $|V_{13}|$ and 
$\langle m_{\nu}\rangle$ are invariant under the change 
of $\a_0$ to $-\a_0$. The quantity $\delta_1$ changes to $-\delta_1$ 
under the exchange of $\a_0$ to $-\a_0$. 
In the following, we confine the region of $\a_0$ to be 
$0\le \a_0 \le \pi$ to discuss above quantities numerically. 

The radiative correction is proportional to 
$\epsilon$, which is a rapidly increasing function of 
$\tan \b$ as seen in Eq.~(8). Therefore, the effect is 
smaller for smaller value of $\tan \b$. In the following, 
we consider two cases, $\tan \b=50$ and $\tan \b=20$. 
For the numerical analysis, we use the experimental data 
given in Eq.~(1). 

\vskip 1mm
\noindent
(a) The angle $\sin^2 2 \theta_{12}$

As we see from Eq.~(26), the solar angle is determined by 
a Majorana phase $\a_0$, the neutrino mass $m_1$ and 
the mixing angle $\sin^2 2 \theta_{12}$ at $M_R$ scale. 
Therefore, when we give the value of $\sin^2 2 \theta_{\odot}$, 
three parameters are constrained and the contour curve 
for a given $\a_0$ is drawn. In Fig.~1, the contour plot of 
$\a_0$ in the $\sin^2 2 \theta_{12}$ and $m_1$ plain 
is shown for $\tan \b=50$ with $\tan^2 \theta_\odot=0.34$. 
The wide values of $\sin^2 2 \theta_{12}$ are allowed 
which may be seen from Eq.~(28). 

A particular feature is that the most of region corresponds 
to $\pi/2\le \a_0\le \pi$. That is, if we choose the value 
of $\a_0$ in this region, almost any value of 
$\sin^2 2 \theta_{12}$ at $M_R$ scale can reproduce the 
experimental solar angle with an appropriate choice of $m_1$, 
which should be greater than, say, 0.02 eV. If the mass 
$m_1<0.01$ eV, the  $\sin^2 2 \theta_{12}$ is stable and 
should reproduce the solar angle precisely at $M_R$ scale.

\vskip 1mm
\noindent
(b) The induced value of $|V_{13}|$

As we see from Eq.~(19), $|V_{13}|$ depends on four 
parameters, $\a_0$, $m_1$, $m_3$ and $\sin^2 2 \theta_{12}$. 
Therefore, we define $|V_{13}|(m_1/m_3)$ and give 
the contour plot in Fig.~2 for $\tan\beta=50$. 
From Fig.~1, we know that the most of the region corresponds to 
$\pi/2\le \a_0\le \pi$. Thus, the point moves to the 
upper right corner in $\sin^2 2 \theta_{12}$ and $m_1$ plain, 
$\a_0$ approaches to $\sim \pi$ from $\pi/2$ and both 
$\sin^2 2 \theta_{12}$ and $m_1$ increase.  Since 
$|V_{13}|(m_1/m_3)$ is proportional to 
$m_1^2\sin 2 \theta_{12} \sin \a_0/2$, its value increases 
rapidly. This situation is seen from Fig.~2 for $\tan \b=50$.
Thus, we may easily expect the value as large as 0.05. 
In order to obtain $|V_{13}|$, we have to multiply $m_3/m_1$ 
which may push its value larger, if $m_3>m_1$.

\vskip 1mm
\noindent
(c) The induced Dirac CP violation phase $\delta$

The induced Dirac CP violation phase contains $\b_0$ which 
is not fixed in this model. Therefore, in general, the Dirac 
phase can take any value, until we fix the value of $\b_0$. 
In order to estimate the Dirac phase aside from $\b_0$, 
we define $\delta_1$ given in Eq.~(29), 
which is obtained by excluding $\b_0$. 
With $\tan \beta=50$, we show in Fig.~3, values of $\sin \delta_1$ 
in the $\sin^2 2 \theta_{12}$ and  $m_1$ plain. 
The solid line shows a curve on which $\sin \delta_1$ takes a 
fixed value. In the right-half domain, the larger value is 
obtained. As we stated, if $\a_0$ enters in the domain 
$-\pi\le\a_0\le0$, the sign of $\sin \delta_1$ changes. 

\vskip 1mm
\noindent
(d) The effective mass of the neutrinoless double beta decay 
$\langle m_\nu\rangle$

The effective mass $\langle m_\nu\rangle$ is proportional 
to $m_1$ as shown in 
Eq.~(30), it becomes larger as $m_1$ increases, while decreases if 
$\sin^2 2 \theta_{12}$ increases. 

All corresponding figures for $\tan \b=20$ are shown in 
Figs.~4, 5, 6. Except for $\langle m_\nu\rangle$, the figures are 
obtained by shifting the larger $m_1$, because the dependence 
of $\tan \b$ is scaled by $\e m_1^2$ as we can see from the 
definition of $h$  in Eq.~(24). The effective mass $\langle m_\nu\rangle$ 
is almost the same as the case of $\tan \b=50$.  

\begin{figure}
\begin{center}
\includegraphics{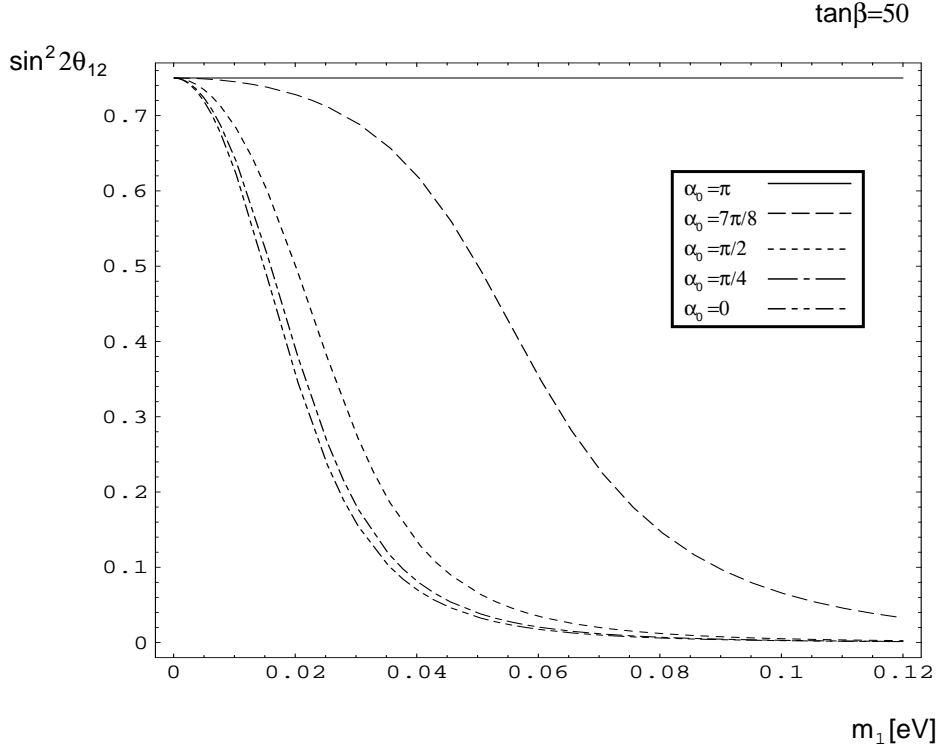}
\caption{Contour plot of $\alpha_0$ in 
$\sin^2 2\theta_{12}$ and $m_1$ plain
to reconstruct the experimental value of 
$\theta_{\odot}$ in the case of $\tan\beta=50$. 
We use as experimental values 
$\tan^2\theta_{\odot}=0.34$($\sin^22\theta_{\odot}\simeq 0.75$),
$\Delta m^2_{\odot}=5\times 10^{-5}[\mathrm{eV}^2]$, 
$\sin^22\theta_{\mathrm{atm}}=1$, and
$\Delta m_{\mathrm{atm}}^2=3\times 10^{-3}[\mathrm{eV}^2]$.
The allowed region is between $\alpha_0=\pi$ and $\alpha_0=0$ curves.}
\end{center}
\end{figure}
\begin{figure}
\begin{center}
\includegraphics{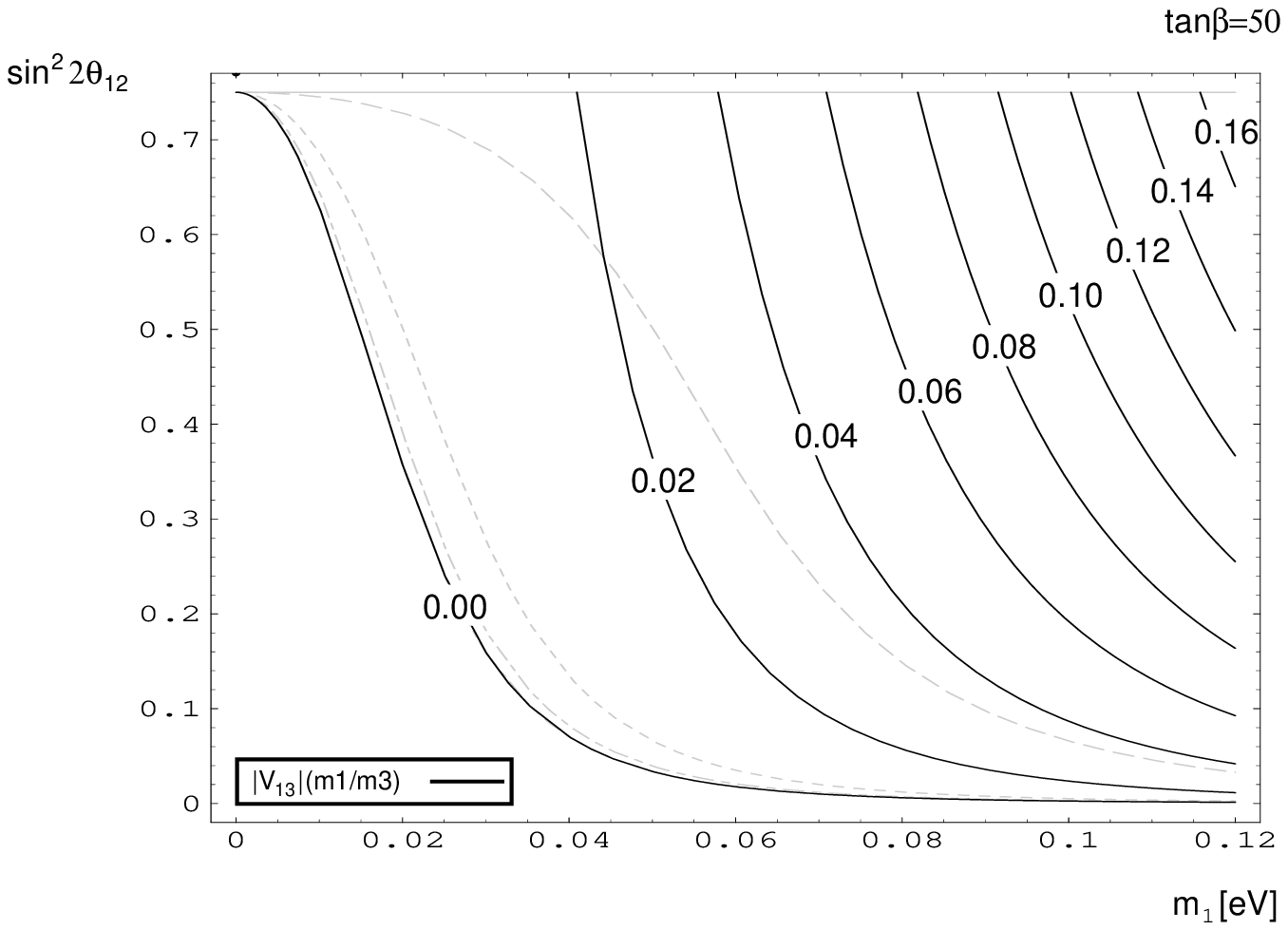}
\caption{Contour plot of $|V_{13}|(m_1/m_3)$ in $\sin^22\theta_{12}$
and $m_1$ plain for $\tan\beta=50$.
We use same values as Fig.~1 for experimental values.
Gray curves show the $\alpha_0$ values as in Fig.~1.}
\end{center}
\end{figure}
\begin{figure}
\begin{center}
\includegraphics{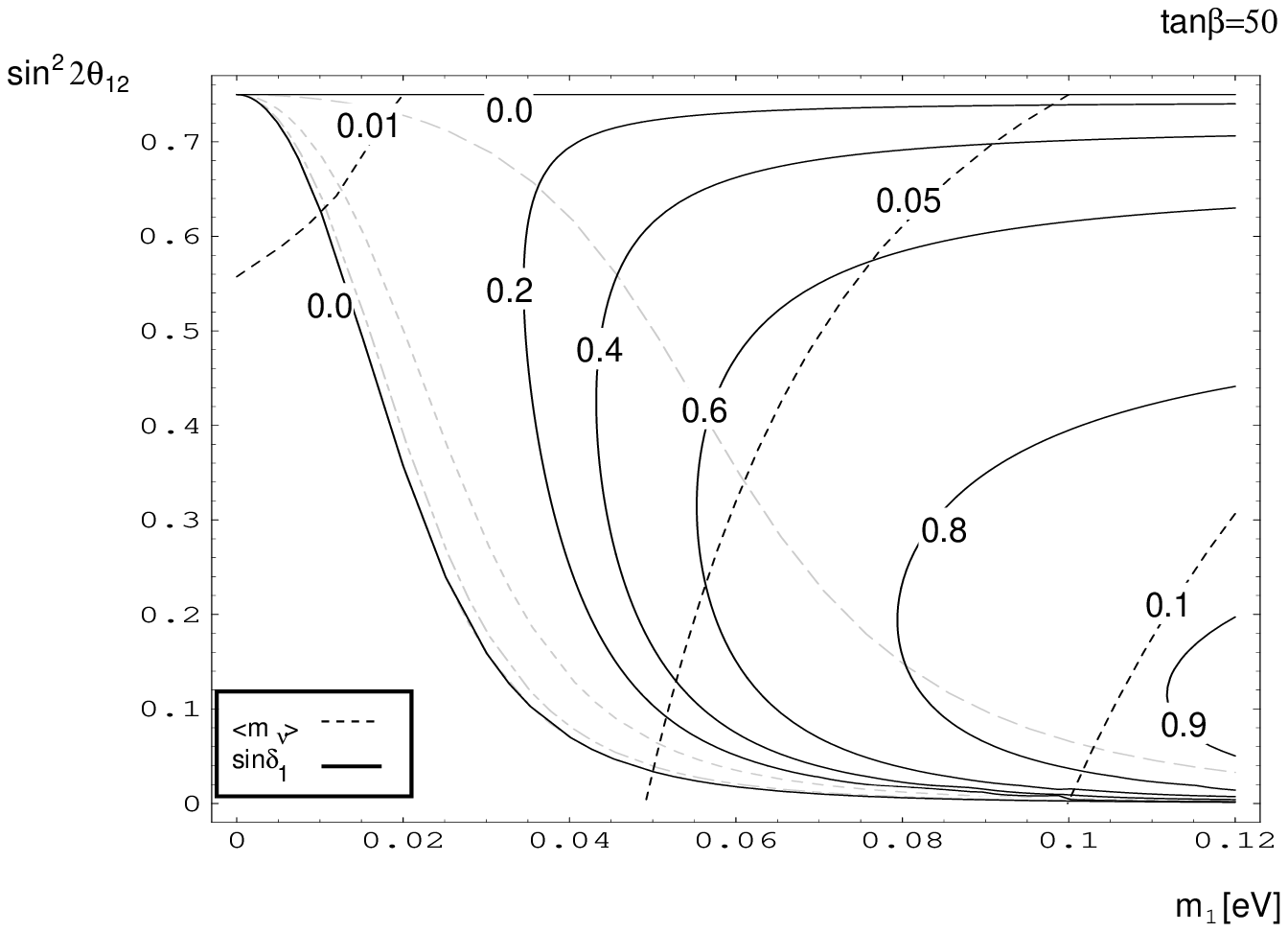}
\caption{Contour plot of $\sin\delta_1$ and $\langle m_{\nu}\rangle$
for $\tan\beta=50$.
We use same values as Fig.~1 for experimental values.
Solid curves denote $\sin\delta_1$ and dashed curves denote 
$\langle m_{\nu}\rangle$. Gray curves show the $\alpha_0$ values
as in Fig.~1.}
\end{center}
\end{figure}
\begin{figure}
\begin{center}
\includegraphics{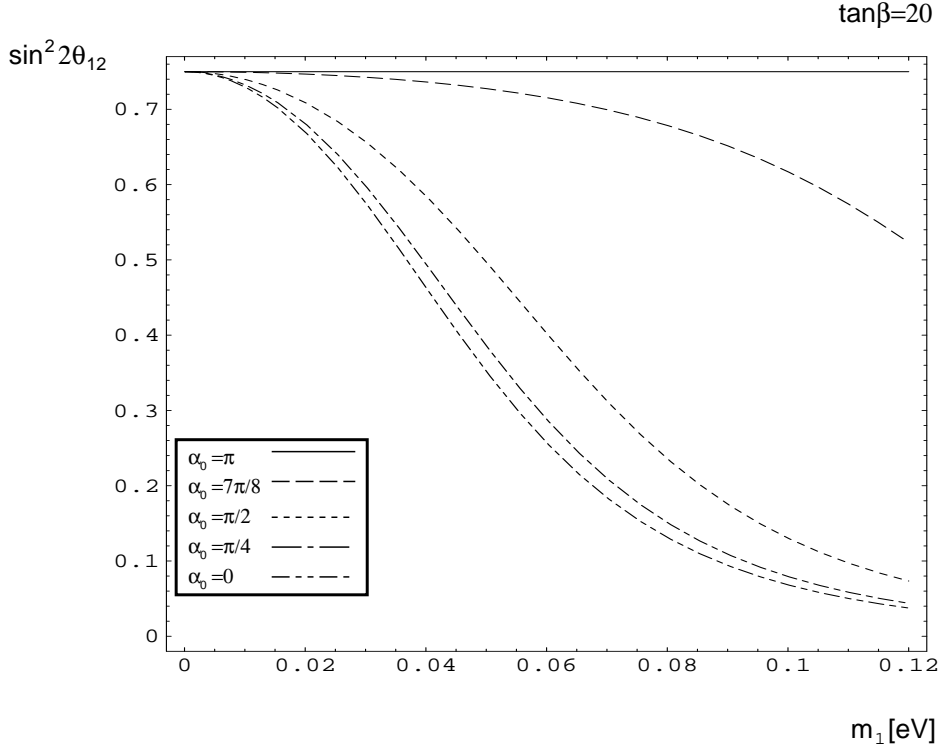}
\caption{Contour plot of $\alpha_0$ in 
$\sin^2 2\theta_{12}$ and $m_1$ plain
to reconstruct the experimental value of 
$\theta_{\odot}$ in the case of $\tan\beta=20$. 
We use same values as Fig.~1 for experimental values.
The allowed region is between $\alpha_0=\pi$ and $\alpha_0=0$ lines.}
\end{center}
\end{figure}
\begin{figure}
\begin{center}
\includegraphics{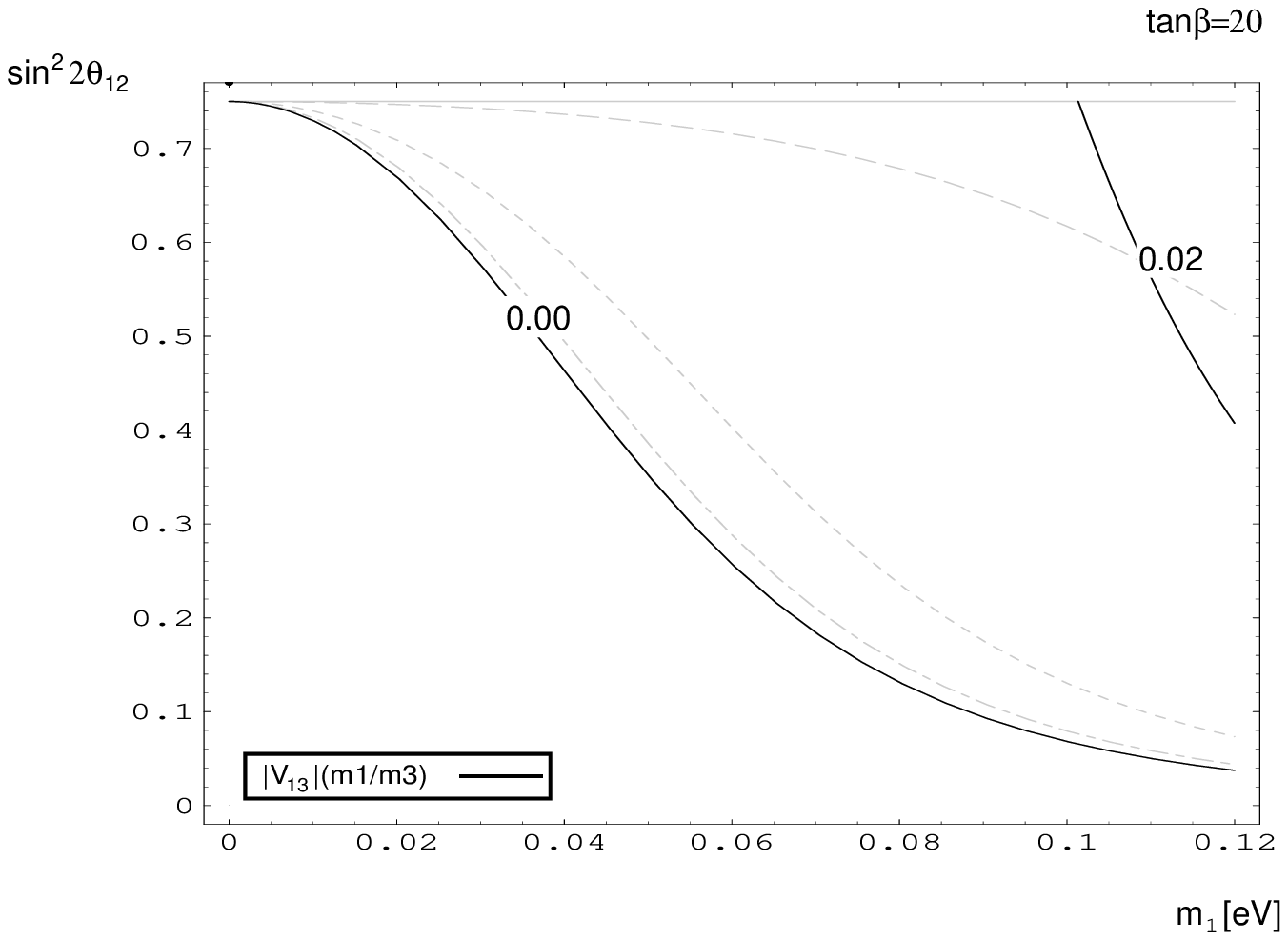}
\caption{Contour plot of $|V_{13}|(m_1/m_3)$ in $\sin^22\theta_{12}$
and $m_1$ plain for $\tan\beta=20$.
We use same values as Fig.~1 for experimental values.
Gray curves show the $\alpha_0$ values as in Fig.~4.}
\end{center}
\end{figure}
\begin{figure}
\begin{center}
\includegraphics{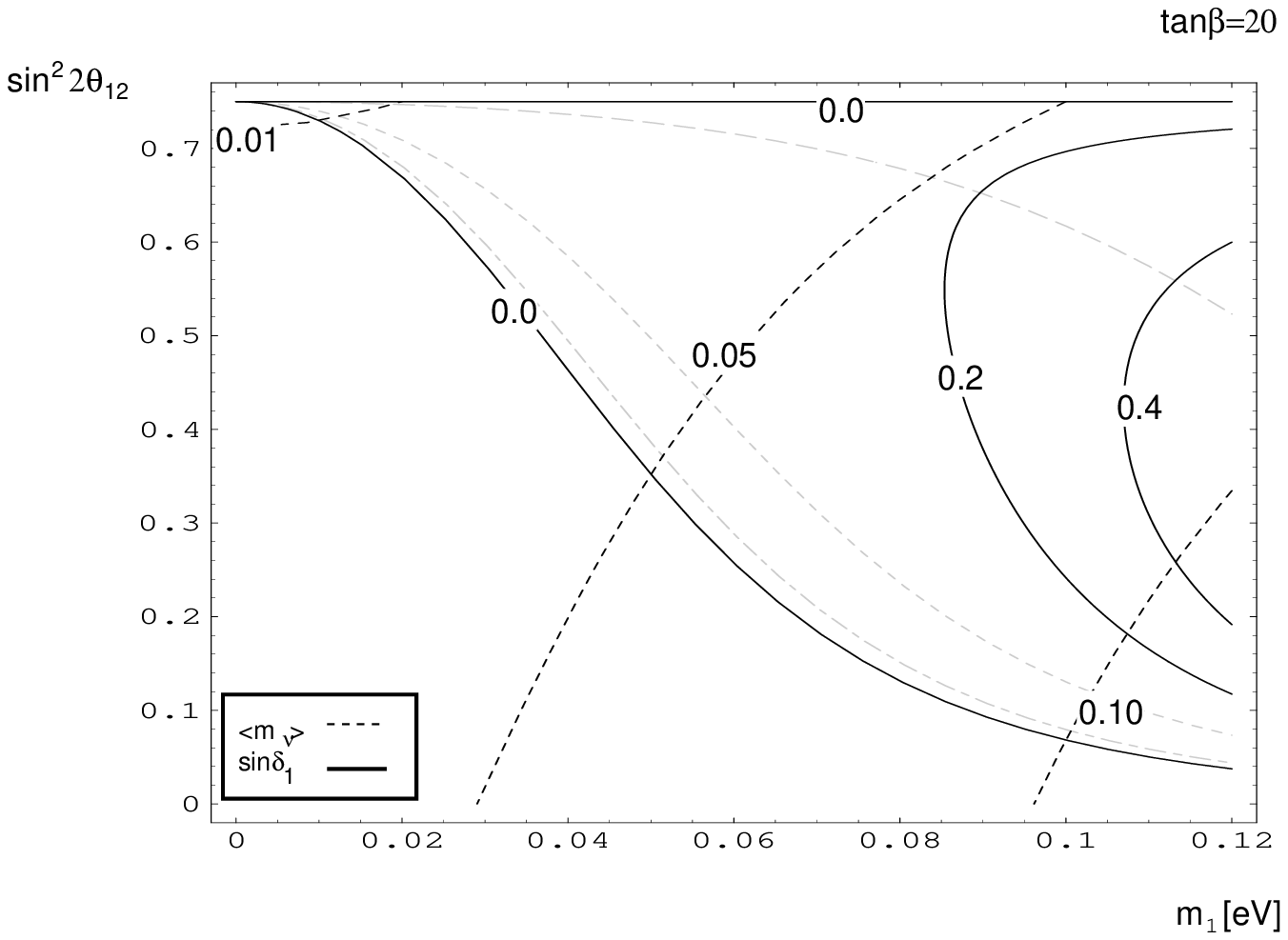}
\caption{Contour plot of $\sin\delta_1$ and $\langle m_{\nu}\rangle$
for $\tan\beta=20$.
We use same values as Fig.~1 for experimental values.
Solid curves denote $\sin\delta_1$ and dashed curves denote 
$\langle m_{\nu}\rangle$. Gray curves show the $\alpha_0$ values
as in Fig.~4.}
\end{center}
\end{figure}

\section{Summary and discussions}

In this paper, we discuss a class of neutrino mass matrix 
which predicts zero or a small value of $|V_{13}|$ and 
found the inequality in Eq.~(28). This constraint gives a 
severe restriction for model building of neutrino mass 
matrix. In particular, the model which predicts a larger 
value of $\tan^2 \theta_{12}$ at $M_R$ scale 
than the experimental value obtained from the solar 
neutrino mixing is excluded. As a result, the bi-maximal 
mixing scheme at $M_R$ scale is excluded, if the experimental 
value $\tan^2 \theta_\odot<1$ is established. 

In this model, $|V_{13}|$ in Eq.~(19) 
at $m_Z$ which is induced radiatively may not be 
small as it is shown in Fig.~2, 
if the neutrino mass $m_1$ is of order 0.05 eV. 
The Dirac phase $\delta_1$ in Eq.~(29) 
at $m_Z$ which is also induced may not be small in 
general as we see in Fig.~3. The effective neutrino mass 
$\langle m_\nu\rangle$ 
in Eq.~(30) is expected to be of order 0.05 eV. All these values for 
$|V_{13}|$, $\delta_1$ and $\langle m_\nu\rangle$  
depend crucially on the mass 
$m_1$ which is assumed to be around 0.05 eV. 

The fact that Majorana phases at $M_R$ scale can induce a 
Dirac phase pushes our dream further to consider the possible 
relation between a Dirac phase which appears in the neutrino 
oscillations and the Majorana phase which appears in the 
leptogenesis. We believe such scenario does exist and the 
finding of the missing link will be the most wonderful and fruitful 
project. 

Recently, Antusch et al.\cite{AKLR} studied the 
quantum effect for the neutrino mass matrix 
which reproduces the Bi-Maximal mixing at the GUT scale, $M_G$. 
They considered the quantum effect due to heavy Majorana 
neutrinos. They considered two cases, (i) the standard model (SM) and (ii) 
the MSSM with $\tan \beta=5$. In both cases, the quantum effect 
from the lightest heavy Majorana neutrino mass, $M_R$ to 
$m_Z$ is very small so that it can be neglected. Therefore, 
they considered a possibility that that 
$\tan^2 \theta_{\rm sol}=1$ at $M_G$  
reduces to the experimental value $\tan^2 \theta_{\rm sol}=0.34$ 
at $M_R$ by the radiative correction. 
They found it possible in a special situation where the Dirac 
mass matrix is in the form of 
$m_D=(v/\sqrt{2}){\rm diag}(1,\e^{\prime},\e^{\prime2})$ with  $\e'<1$, which  in 
turn means that $M_R\sim 10^{14}$GeV. 
This model has two special features: One is that the  Dirac mass 
matrix has the inverse mass hierarchy which disagrees with 
the naive expectation from the GUT scenario. The other is that 
the scale of $M_R$ is larger than the ordinary expectation, 
$M_R\sim 10^{13}$GeV in order to have larger Yukawa coupling 
constants, $y_\nu$, related to the Dirac neutrino mass. 
In our case, we consider the large $\tan \beta$ case so that 
Yukawa coupling constants $y_\nu$ are small. Thus, 
the correction is negligible and our result is valid even at 
$M_G$.

\vskip 5mm
{\Huge Acknowledgment} 
This work is supported in part by 
the Japanese Grant-in-Aid for Scientific Research of
Ministry of Education, Science, Sports and Culture, 
No.12047218.
 
\newpage

\setcounter{section}{0}
\renewcommand{\thesection}{\Alph{section}}
\renewcommand{\theequation}{\thesection .\arabic{equation}}
\newcommand{\apsc}[1]{\stepcounter{section}\noindent
\setcounter{equation}{0}{\Large{\bf{Appendix\,\thesection:\,{#1}}}}}

\apsc{Diagonalization of the neutrino mass matrix}

We define ${\rm diag}(1,1,\a)O=OX$,
\begin{eqnarray}
X &=&1-\e O^T{\rm diag}(0,0,1)O\nonumber\\
  &=&1-\e \begin{pmatrix} s_{12}^2s_{23}^2&
   s_{12} c_{12}s_{23}^2 &
     s_{12}s_{23}c_{23}\cr
     s_{12} c_{12}s_{23}^2 & c_{12}^2s_{23}^2
    & c_{12}s_{23}c_{23}\cr
      s_{12}s_{23}c_{23}&  c_{12}s_{23}c_{23}&c_{23}^2 \cr\end{pmatrix}\;,
\end{eqnarray}
where $\e=1-\a$ is a small positive quantity and its value is 
given in Eq.~(8), then 
we consider the mass matrix transformed by $O$ as
\begin{eqnarray}
\bar m_{\nu} \equiv O^Tm_\nu(m_Z) O=X D_\nu X^T\;.
\end{eqnarray}
Now, we diagonalize  $\bar m_{\nu}$. 

In order to diagonalize this matrix directly, we consider
the Hermite matrix $\bar m_\nu^\dagger \bar m_\nu$ 
\begin{eqnarray}
\bar m_{\nu}^\dagger \bar m_{\nu}
\simeq(1-4\e  s_{12}^2s_{23}^2)M_1^2 +Y\;,
\end{eqnarray}
where elements of $Y$ are given up to the 1st order of $\e$ as 
\begin{eqnarray}
Y_{11}&=& 0\;,\;Y_{22}=\tilde \Delta_{21}\;,\;
Y_{33}=\tilde \Delta_{31}\;,\nonumber\\
Y_{12}&=& Y_{21}^*
 =-2\e  s_{12} c_{12}s_{23}^2(M_1^2+M_2^2+2M_1M_2
 e^{i\a_0})M_1^2\;,\nonumber\\
Y_{13}&=& Y_{31}^*
 =-\e  s_{12}s_{23}c_{23}(M_1^2+M_3^2+2M_1M_3e^{i\b_0})\;,\nonumber\\
Y_{23}&=&Y_{32}^*
= -\e  c_{12}s_{23}c_{23}(M_2^2+M_3^2+2M_2M_3 
  e^{i(\b_0-\a_0)})\;.\nonumber\\
\end{eqnarray}
\begin{eqnarray}
\tilde \Delta_{21} &=&\Delta_{21}
   -4\e s_{23}^2( c_{12}^2M_1^2 - s_{12}^2M_2^2)\;,
\nonumber\\
\tilde \Delta_{31}&=&\Delta_{31}
   -4\e ( c_{23}^2M_3^2- s_{12}^2s_{23}^2M_1^2)\;.
\end{eqnarray}
Since $\bar m_{\nu}^\dagger \bar m_{\nu}$ is an Hermite 
matrix, it is diagonalized by the unitary transformation as 
$V^\dagger \bar m_{\nu}^\dagger \bar m_{\nu}V
\simeq(1-4\e  s_{12}^2s_{23}^2)M_1^2 +V^\dagger Y V$. 
The diagonalization of the matrix $Y$ can be achieved by using the 
see-saw technique, because $|Y_{33}|$ is much larger than 
all other terms. That is, by using the unitary matrix 
\begin{eqnarray}
V_{3} \simeq \begin{pmatrix}1&0&
        \frac{Y_{13}}{Y_{33}} \cr
    0&1&\frac{Y_{23}}{Y_{33}}\cr
    -\frac{Y_{13}^*}{Y_{33}}    &
    -\frac{Y_{23}^*}{Y_{33}}   &1\cr\end{pmatrix}\;.
\end{eqnarray}
$Y$ is block diagonalized 
in a good accuracy as 
\begin{eqnarray}
V_3^\dagger Y V_3
&\simeq& 
\begin{pmatrix}-\frac{2|Y_{13}|^2}{Y_{33}} 
  &Y_{12}-\frac{2Y_{13}Y_{23}^*}{Y_{33}}&0 \cr
  Y_{12}^*-\frac{2Y_{13}^*Y_{23}}{Y_{33}}
  &Y_{22}-\frac{2|Y_{23}|^2}{Y_{33}} &0 \cr
        0&0&Y_{33}+\frac{|Y_{13}|^2+|Y_{23}|^2}{Y_{33}}\cr\end{pmatrix}
        \nonumber\\
        &\simeq &
        \begin{pmatrix}0 &Y_{12}&0 \cr
  Y_{12}^* &Y_{22} &0 \cr
        0&0&Y_{33}\cr\end{pmatrix}
        \;,
\end{eqnarray}
where in the last equation, we neglected 
the see-saw induced terms because 
$|Y_{13}/Y_{33}|$ and $|Y_{23}/Y_{33}|$ are 
much smaller than 1 and $Y_{13}$ and $Y_{23}$ are 
same order of $Y_{12}$ and $Y_{22}$.

In the following, we use $M_1=M_2$ for 
terms proportional to $\e$ to simplify the expression.  
The matrix in Eq.~(A7) is diagonalized by 
\begin{eqnarray}
V_{12}=\begin{pmatrix}c & -s e^{i\a_0/2} &0\cr
              s e^{-i\a_0/2} &c &0\cr
              0&0&1\cr \end{pmatrix}\;,
\end{eqnarray}
with $c=\cos \theta$ and $s=\sin \theta$ as
\begin{eqnarray}
V_{12}^\dagger (V_{3}^\dagger Y V_{3})V_{12}=
{\rm diag}(\lambda_1,\lambda_2 ,\lambda_3)\;,
\end{eqnarray}
where
\begin{eqnarray}
\tan 2\theta= \frac{4\e M_1^2 \sin 2\theta_{12} s_{23}^2 
\cos \frac{\a_0}{2}}{\tilde \Delta_{21}}\;,
\end{eqnarray}
and 
\begin{eqnarray}
\lambda_1\simeq \frac{\tilde \Delta_{21}}2
 \left(1-\frac1{\cos 2\theta}\right)\;,\;\;
\lambda_2\simeq \frac{\tilde \Delta_{21}}2
\left(1+\frac1{\cos 2\theta}\right)\;,\;\;
\lambda_3\simeq \tilde \Delta_{31}\;.
\end{eqnarray}
Neutrino masses at $m_Z$ are obtained by 
$m_i^2= (1-4\e s_{12}^2s_{23}^2)M_1^2 +\lambda_i$ and 
\begin{eqnarray}
\Delta m_{\odot}^2&\equiv& m_2^2-m_1^2=
  \frac{\tilde \Delta_{21}}{\cos 2\theta}\;,
  \nonumber\\
\Delta m_{\rm atm}^2&\equiv& 
m_3^2-m_2^2\simeq m_3^2-m_1^2\simeq \tilde \Delta_{31}
\;.
\end{eqnarray}
We find that the angle 
$\theta$ is expressed by
\begin{eqnarray}
\sin 2\theta= \frac{4\e M_1^2 \sin 2 \theta_{12}s_{23}^2\cos 
\frac{\a_0}{2} }{\Delta m_{\odot}^2} 
\;.
\end{eqnarray}
By taking into account of 
\begin{eqnarray}
(OV_{3}V_{12})^T m_\nu(m_Z) OV_{3}V_{12}
 \simeq {\rm diag}(m_1,m_2 e^{i\a_0},m_3 e^{i\b_0})\;,
\end{eqnarray}
with $m_i>0$, we find the mixing matrix $V$ which satisfies 
$V^T m_\nu(m_Z)V={\rm diag}(m_1,m_2,m_3)$ is
$V\equiv OV_{3}V_{12}{\rm diag}(1,e^{-i\a_0/2},e^{-i\b_0/2})$ is given
by 
\begin{eqnarray}
V\simeq
\begin{pmatrix} c_{12}c- s_{12}s  e^{-i\a_0/2}
&-( s_{12}c+ c_{12}s  e^{i\a_0/2})& -|V_{13}|e^{i\rho}\cr
c_{23}( s_{12}c+  c_{12}s  e^{-i\a_0/2})& 
c_{23}( c_{12}c- s_{12}s  e^{i\a_0/2})& -s_{23}\cr
s_{23}( s_{12}c+  c_{12}s  e^{-i\a_0/2})& 
s_{23}( c_{12}c- s_{12}s  e^{i\a_0/2})& c_{23}\cr\end{pmatrix}
\begin{pmatrix}1&0&0\cr 0&e^{-i\a_0/2}&0\cr 0&0&e^{-i\b_0/2}
\cr\end{pmatrix}\;,\nonumber\\
\end{eqnarray}
where 
\begin{eqnarray}
|V_{13}|&=&\frac{\epsilon m_1m_3 \sin 2 \theta_{12}\sin 2\theta_{23}
\sin\frac{\alpha_0}{2}}
{\tilde \Delta m^2_{31}}\;,\nonumber\\
\rho &=& \frac{\pi}2-\frac{\a_0}2+\b_0\;.
\end{eqnarray}

\end{document}